\documentclass[10pt,pra,twocolumn,showpacs,floatfix]{revtex4}

\usepackage{graphicx}
\usepackage{amssymb}
\usepackage{times}
\usepackage{amsmath}
\usepackage{amsthm}
\usepackage{subfigure}
\usepackage{float}
\usepackage[breaklinks]{hyperref}

\input epsf.tex
\usepackage{graphicx}
\usepackage{amsthm}

\begin{document}

\title{Difference Weak Measurement}

\author{Jing-Zheng Huang\footnote{Email address: jzhuang1983@sjtu.edu.cn}, Chen Fang, and Guihua Zeng}
\affiliation
{State Key Laboratory of Advanced Optical Communication Systems and Networks, Center of Quantum Sensing and Information Processing, Shanghai Jiao Tong University, Shanghai 200240, China}


\begin{abstract}
We propose the difference weak measurement scheme, and illustrate its advantages for measuring small longitude phase-shift in high precision. Compared to the standard interferometry and standard weak measurement schemes, the proposed scheme has much higher resolution in present of various practical imperfections, such as alignment error and light intensity variation error. Moreover, we highlight the advantage of utilizing complex weak value, where its imaginary part can reduce the harmful effect induced by channel decoherence. Finally, we propose closed-loop scenario to solve the narrow dynamic range problem obsessing the current weak measurement schemes. Difference weak measurement scheme simultaneously fulfills the requirements of high precision, wide dynamic range and strong robustness, which makes it a powerfully practical tool for phase-shift measurement and other metrological tasks.
\end{abstract}

\maketitle

\section{Introduction}
Among those proposals for precision metrology, weak value amplification (WVA)\cite{Aharonov1988} technique has attracted much attention. It has been proven in theory\cite{Feizpour2011,Jordan2014} and demonstrated in experiments\cite{Viza2015,Alves2017} that unexpected effects of certain kinds of practical imperfections, such as 1/f noise and alignment errors, can be efficiently suppressed. Especially in Ref.\cite{Brunner2010}, the authors showed that a small phase-shifting can be transferred to an amplified average frequency shifting through WVA, which outperforms the standard interferometry technique if the misalignment errors are taken into account. Although this method has been experimentally demonstrated\cite{Xu2013,Fang2016}, the requirement of high resolution spectrum analysis increases the detection complexity. Moreover, the current weak measurement schemes suffer from the disadvantage of narrow dynamic range: the basic principle of weak measurement restricts the maximum measurable value of parameter at a level much smaller than 1.

In this work, we propose a new scheme called difference weak measurement (DWM). In contrast to the previous weak measurement schemes, DWM applies the WVA technique to amplify the phase-shift before detection, and reveals the signal information via light intensity detection as well as the standard interferometry, thus the complex spectrum analysis is avoided. For this reason, DWM can be implemented by closed-loop scenario, which solves the narrow dynamic range problem. Moreover, instead of choosing pure real or pure imaginary WVA factor as in the previous proposals\cite{Brunner2010,Xu2013,Fang2016,Strubi2013,Qiu2017,Hu2017}, our scheme allows setting complex weak value, thus simultaneously adopts the advantages of having signal-to-noise ratio against technical noise (with help of the real part of the weak value) and remaining high precision in the present of decoherence (with help of the imaginary part of the weak value).

Moreover, we apply DWM to realize high precision phase-shift measurement, which is at the core of many modern metrological applications including but not limited by optic coherence tomography\cite{Schmitt1999}, interferometric surface profilers\cite{Bowe1998}, optic gyro\cite{Lefevre2014}, gravity wave detection\cite{Abbott2016} and so on. In theory, the minimum measurable phase-shift is limited by the shot noise limit, which would be rapidly decreased along with the increasing measured events. However in practice, the technical noises, such as the relative intensity noise and dark current noise inside the detector, induce measurement uncertainty that usually much higher than the theoretical limit. Our analysis shows that DWM scheme addresses these problems, and takes advantages over standard interferometry and previous proposed weak measurement schemes under practical imperfections.

This paper is organized as follows: 
In Sec.II we propose the theoretical model of DWM, and present it in a phase-shift measurement scheme. In Sec.III we analyze the robustness of DWM-based phase-shift scheme against practical imperfections, such as alignment errors, detection errors and decoherence, and compare it to the previous proposals. In Sec.IV, we extend the dynamic range of DWM scheme by introducing the closed-loop scenario. Finally, brief conclusion of this paper is made in Sec.V.

\section{Difference Weak Measurement}
To build up the theoretical model of difference weak measurement, we begin with considering a process of measuring an unknown parameter $g$ through an unitary operation, i.e.,
$$\hat{U} = \exp[ig\hat{A}],$$
where $\hat{A}$ represents a Pauli matrix\cite{Nielsen2000}, with eigenvalues of -1 and 1, acting on a two-level state, i.e., $|\phi_i\rangle = (|0\rangle + |1\rangle)/\sqrt{2}$, with $|0\rangle$ and $|1\rangle$ the eigenvectors of $\hat{A}$. After the operation, $|\phi_i\rangle$ evolves to be
$$
|\phi_{out}\rangle = e^{ig}|0\rangle+e^{-ig}|1\rangle.
$$
The information of $g$ can be extracted by performing projective measurement on the state $\phi_{out}$. Specifically, by projecting it to a state $|\phi_f\rangle$, the final state before being measured is:
\begin{equation}\label{eq:ig}
|\phi_f\rangle\langle\phi_f|\phi_{out}\rangle = \langle\phi_f|\phi_i\rangle e^{iA_wg}|\phi_f\rangle,
\end{equation}
where $A_w=\langle\phi_f|\hat{A}|\phi_i\rangle/\langle\phi_f|\phi_i\rangle$ is the so-called "weak value"\cite{Aharonov1988}, which can be anomalous for realizing weak value amplification. For instance, by setting $|\phi_f\rangle = \cos(-\pi/4+\gamma)|0\rangle + \sin(-\pi/4+\gamma)|1\rangle$ with $\gamma \ll 1$, we get 
$$A_w \approx 1/\gamma \gg 1.$$ 
Therefore the phase-shift is effectively amplified, however the global phase can not be directly measured, unless an additional pointer is included.

We are now ready to introduce the difference weak measurement scheme, by involving an "ancillary system" state $|\phi_i\rangle_A$ and a "pointer" state $|\psi_i\rangle_B$, with both of two-level degree of freedom. The interaction between the system and the pointer is described by an unitary operator:
\begin{equation}
\hat{U}_{int} = \exp[ig\hat{A}\hat{B}] = \cos g(\hat{I}_A\otimes\hat{I}_B) + i\sin g(\hat{A}\otimes\hat{B}).
\end{equation}
where $\hat{A}$ and $\hat{B}$ are Pauli matrices\cite{Nielsen2000} with eigenvalues of 1 and -1, which act on the ancillary system and pointer respectively; $g$ is the parameter of interest, which indicates the coupling strength and satisfies $g \ll 1$; $\hat{I}_A$ and $\hat{I}_B$ are identity matrices. The last expression is based on the fact that $\hat{A}^2=\hat{B}^2=\hat{I}$\cite{Nielsen2000}.

By post-selecting the system to state $|\phi_f\rangle_A$ that satisfying $\langle\phi_f|\phi_i\rangle_A \ll 1$, the pointer state will be collapsed to (unnormalized):
\begin{equation}
\begin{array}{lll}
|\psi_f\rangle_B &= \langle\phi_f|\hat{U}_{int}|\phi_i\rangle_A|\psi_i\rangle_B \\
                 &= \langle\phi_f|\phi_i\rangle_A[\cos g\hat{I}_B + i\sin g\cdot A_w\cdot\hat{B}]|\psi_i\rangle_B\\
                 &\approx \langle\phi_f|\phi_i\rangle_A\exp[i (A_wg)\hat{B}]|\psi_i\rangle_B,
\end{array}
\end{equation} 
where the last approximation is established based on the assumption of $A_wg \ll 1$. To extra the parameter information, a projective measurement on the pointer is performed. If we set $|\psi_i\rangle_B = |0\rangle_B + |1\rangle_B$ with $|0\rangle_B$ and $|1\rangle_B$ being the basis of pointer, the final (normalized) state of the pointer becomes:
\begin{equation}\label{eq:iAg}
|\psi_f\rangle_B = e^{iA_wg}|0\rangle + e^{-iA_wg}|1\rangle.
\end{equation}
Instead of being a global phase in Eq.(\ref{eq:ig}), the one appears in Eq.(\ref{eq:iAg}) is a relative phase, thus can be directly measured.

\begin{figure}[!h]
	\centering
	\subfigure[]{
		\begin{minipage}{9cm}
			\centering
			\includegraphics[width=0.9\textwidth]{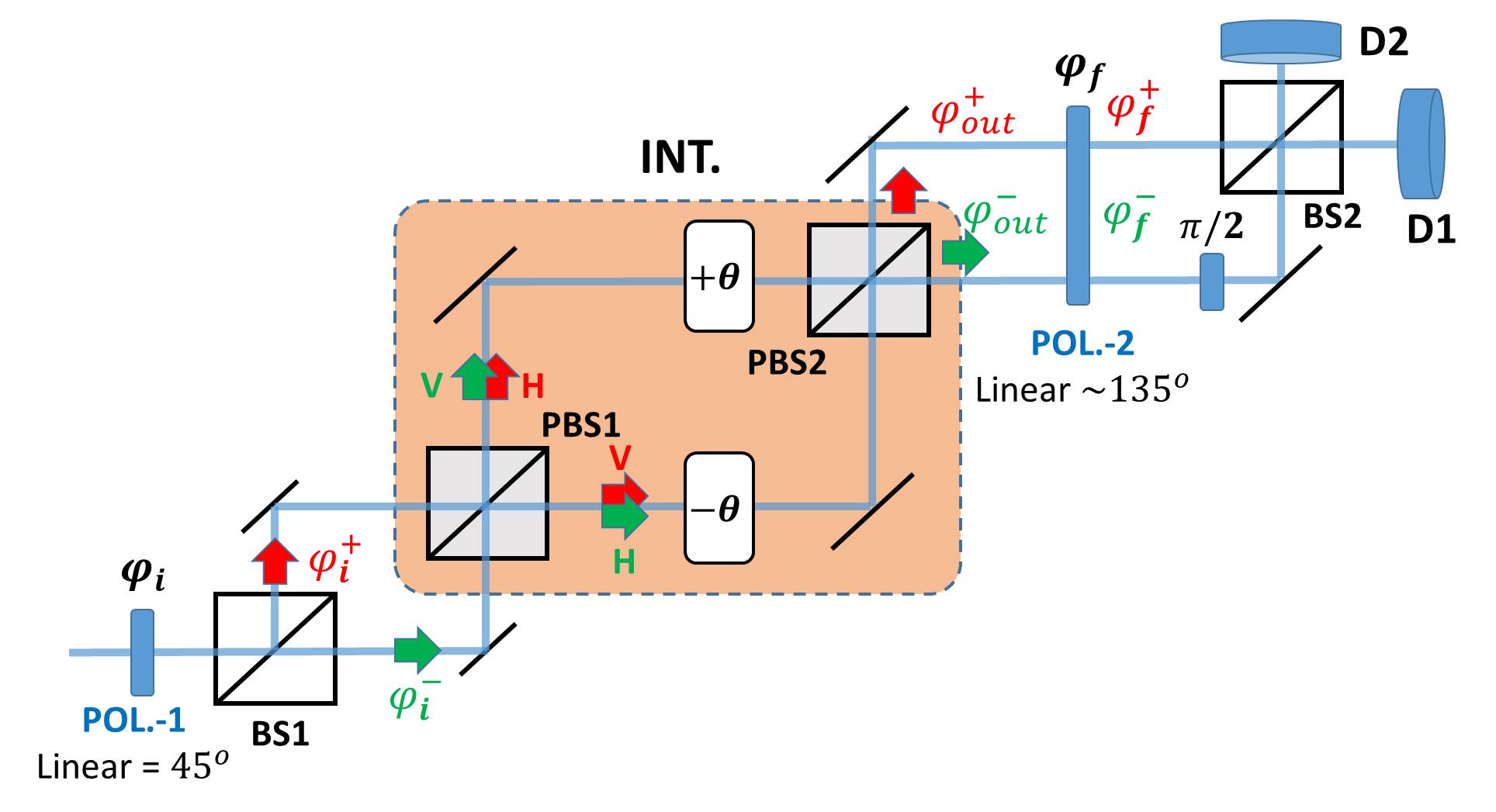}
		\end{minipage}
	}
	\subfigure[]{
		\begin{minipage}{6cm}
			\centering
			\includegraphics[width=0.9\textwidth]{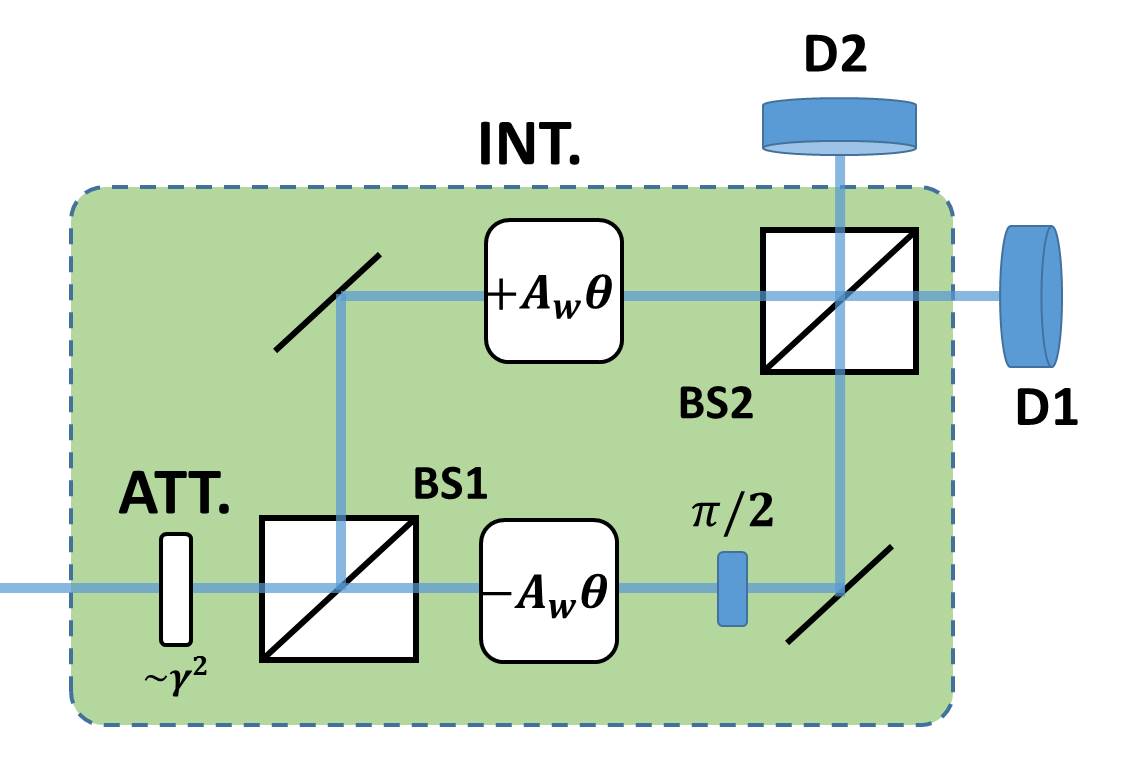}
		\end{minipage}
	}
	\caption{(Color online). (a) The schematic paradigm of DWM-based phase-shift measurement scheme. (b) The equivalent interferometric scheme. $\pm\theta$: the to-be-measured phase-shift; POL: linear polarizer; PBS: polarizing beam splitter; BS: beam splitter; ATT: intensity attenuation; D: photo-detector.}
	\label{fig_scheme_open}
\end{figure}

For clarity, we present a phase-shift measurement experimental proposal to illustrate the DWM scheme.  
Note that although we illustrate the idea through a polarization-based Mach-Zehnder interferometry, this method can be easily applied to other types of interferometer, such as Michelson or Sagnac interferometers. The DWM process comprises of four basic stages: pre-selection, interaction, post-selection and detection. To be specific, we consider a DWM scheme depicted in Fig.\ref{fig_scheme_open}, the goal of this scheme is to measure an unknown phase-shift $\theta$ introduced by different path lengths. The four basic stages are described as follows:

(1) \textit{The pre-selection stage}: an incoming light beam with central wavelength of $\lambda_0$ is linearly polarized to the pre-selected state:
$$|\varphi_i\rangle = \frac{1}{\sqrt{2}}(|H\rangle + |V\rangle)$$ 
by a polarizer, where H and V stand for polarizations in horizontal and vertical directions respectively. 

(2) \textit{The interaction stage}: in this stage, the light is split into two paths, denoted as '$+$' (red, upper path) and '$-$' (green, lower path), by a beam splitter (BS1) before entering to the polarizing beam splitter (PBS1):
$$|\varphi_i\rangle \rightarrow \frac{1}{\sqrt{2}}(i|\varphi_{+}\rangle + |\varphi_{-}\rangle).$$
After entering PBS1, a phase-shift of $\theta$ is then introduced between two orthogonal polarization components, the polarization state of the $\varphi_{\pm}$ evolves to 
$$|\varphi^{\pm}_{out}\rangle = \frac{1}{\sqrt{2}} (e^{\pm i\theta}|H\rangle + e^{\mp i\theta}|V\rangle).$$
This process can be described by unitary operator $U_{\pm}(\theta) = \exp[\pm i\theta\hat{A}]$, where $\hat{A} \equiv |H\rangle\langle H|-|V\rangle\langle V|$. After recombining by the second PBS (PBS2), $|\varphi^{+}_{out}\rangle$ and $|\varphi^{-}_{out}\rangle$ come out from the upper port and lower port of PBS2 respectively.

(3) \textit{The post-selection stage}: the output lights are post-selected in the same polarization state $$|\varphi_f\rangle = \cos\alpha|H\rangle + \sin\alpha|V\rangle,$$ 
where $\alpha=-\pi/4 + \gamma/2$ with $\gamma \ll 1$. After the post-selection, lights propagate in two paths are in the same polarization state with different relative phase-shift: 
\begin{equation}
\begin{array}{lll}\label{eq:final_state}
|\varphi_f^{\pm}\rangle &= \langle\varphi_f|\varphi^{\pm}_{out}\rangle = \frac{\langle\varphi_f|\varphi_i\rangle}{\sqrt{2}} (\cos\theta + iA_w\sin\theta)\\
&= \frac{\langle\varphi_f|\varphi_i\rangle}{\sqrt{2}} N'e^{i\theta'}\\
&\approx \frac{\langle\varphi_f|\varphi_i\rangle}{\sqrt{2}} e^{\pm iA_w\theta}|\varphi_f\rangle,
\end{array}
\end{equation}
where  $A_w \equiv \frac{\langle\varphi_f|\hat{A}|\varphi_i\rangle}{\langle\varphi_f|\varphi_i\rangle} \simeq \frac{1}{\gamma}$, $N' = \sqrt{\cos^2\theta+\sin^2\theta A_w^2}$, and 
$$\theta' = \tan^{-1}(A_w\tan\theta) \simeq A_w\theta.$$ 
The last approximation is established when $A_w\theta \ll 1$. Note that in the current setting of $\varphi_i$ and $\varphi_f$, $A_w$ is pure real.

(4) \textit{The detection stage}: In the last stage, we extract the phase-shift information by observing the interference between $\varphi_f^{+}$ and $\varphi_f^{-}$. Two photo-detectors (denoted as $D1$ and $D2$) are placed at one of the output port, the total probabilities of detecting light by these detectors are
\begin{equation}\label{eq:pd}
\begin{array}{lll}
P_{D1} \simeq \frac{\gamma^2}{2} [1 + \sin(2A_w\theta)],\\
P_{D2} \simeq \frac{\gamma^2}{2} [1 - \sin(2A_w\theta)].
\end{array}
\end{equation}
Assume that the input light intensity is $I_0$, the detected light intensity of $D1$ and $D2$ are given by $I_{D(1,2)} = P_{D(1,2)}I_{D(1,2)}$, and the output signal is given by the subtraction of light intensities  detected by D1 and D2 (denoted as $I_{D1}$ and $I_{D2}$):
\begin{equation}\label{eq:dwm}
\begin{array}{lll}
I_{S(DWM)} &= I_{D1} - I_{D2} = I_0(P_{D1} - P_{D2}) \\
           &\approx 2\gamma^2 I_0 A_w\theta,
\end{array}
\end{equation}

\begin{figure}[!h]
	\centering
	\subfigure[]{
		\begin{minipage}{7cm}
			\centering
			\includegraphics[width=0.9\textwidth]{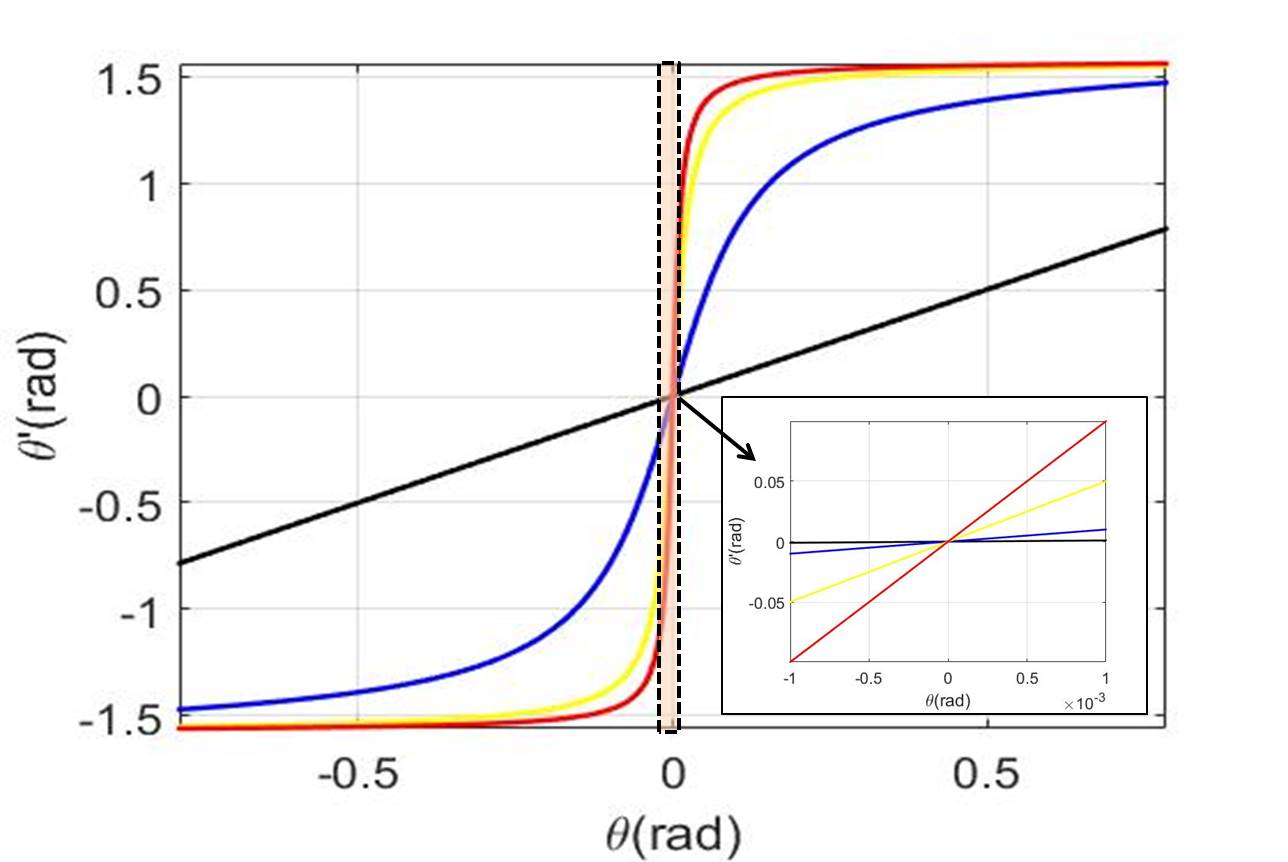}
		\end{minipage}
	}
	\subfigure[]{
		\begin{minipage}{8cm}
			\centering
			\includegraphics[width=0.85\textwidth]{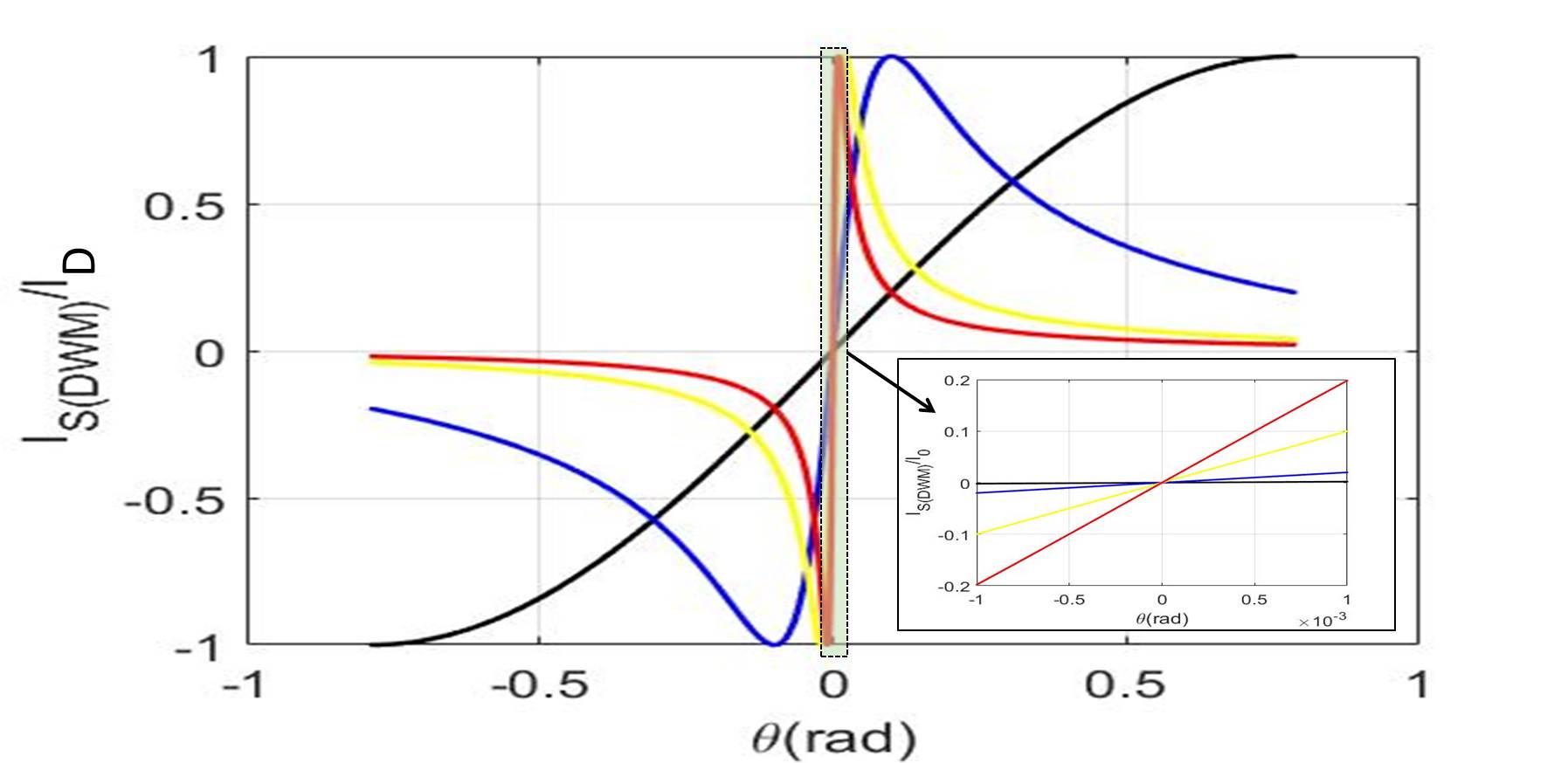}
		\end{minipage}
	}
	\caption{(Color online). The effect of phase amplification. (a) The relation between original phase-shift $\theta$ and amplified phase-shift $\theta'$; (b) The relation between $\theta$ and the relative signal intensity $I_{S(DWM)}/I_D$, where $I_D = I_{D1} + I_{D2}$ is the total detected light intensity. The different colors correspond to different $A_w$ values: 1 (black), 10 (blue), 50 (yellow), and 100(red).}
	\label{fig_curve}
\end{figure}

For simplicity, here we assume the efficiencies of both detectors are 1. In effect, the (open-loop) DWM scheme is equivalent to a standard interferometry with amplitude attenuated by a factor of $\gamma$ and phase-shift amplified by a factor of $A_w \approx 1/\gamma$, as is shown in Fig.\ref{fig_scheme_open}(b). The variation of effective amplified phase $\theta'$ and output signal $I_{S(DWM)}$ along with the phase-shift $\theta$ under different weak values are shown in Fig.\ref{fig_curve}.
%
%
\begin{figure}[!h]
	\centering
	\subfigure[]{
		\begin{minipage}{8cm}
			\centering
			\includegraphics[width=0.9\textwidth]{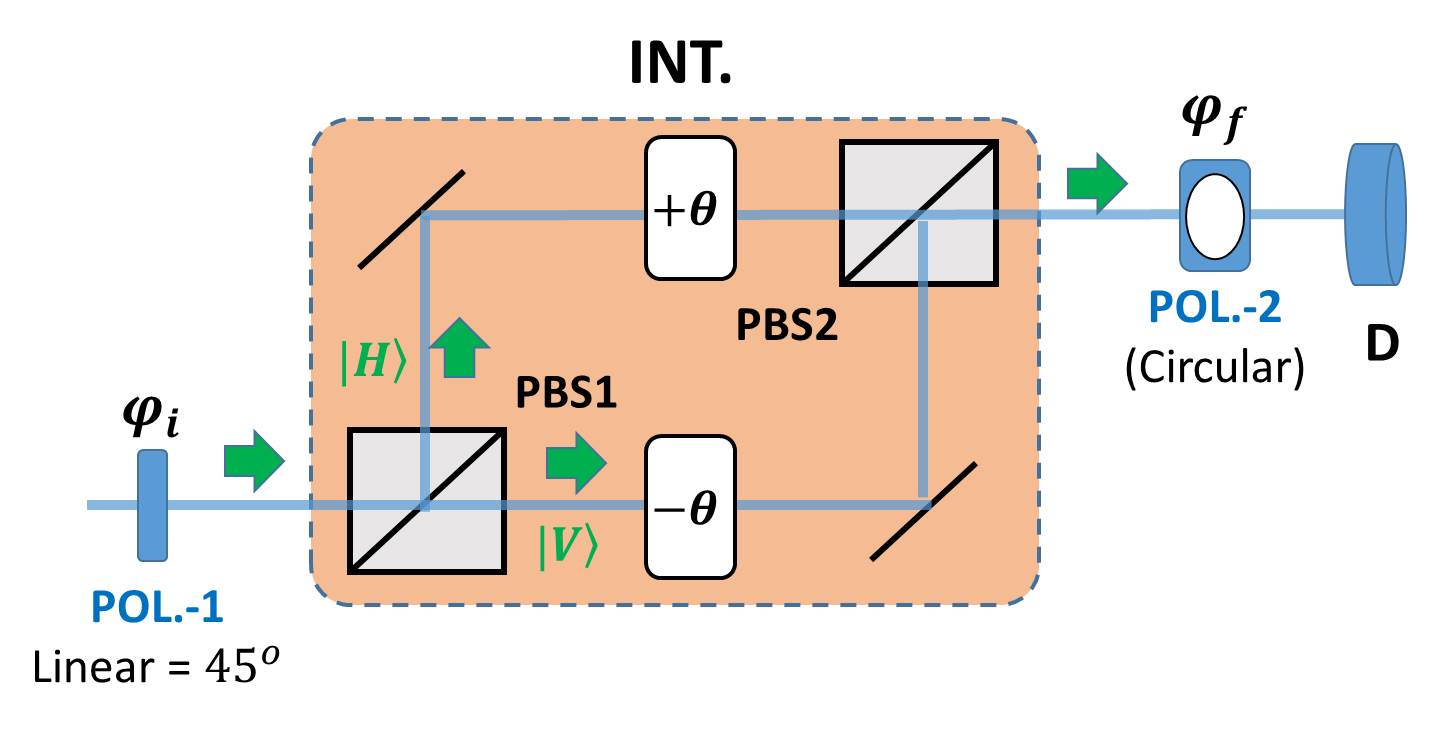}
		\end{minipage}
	}
	\subfigure[]{
		\begin{minipage}{9cm}
			\centering
			\includegraphics[width=0.9\textwidth]{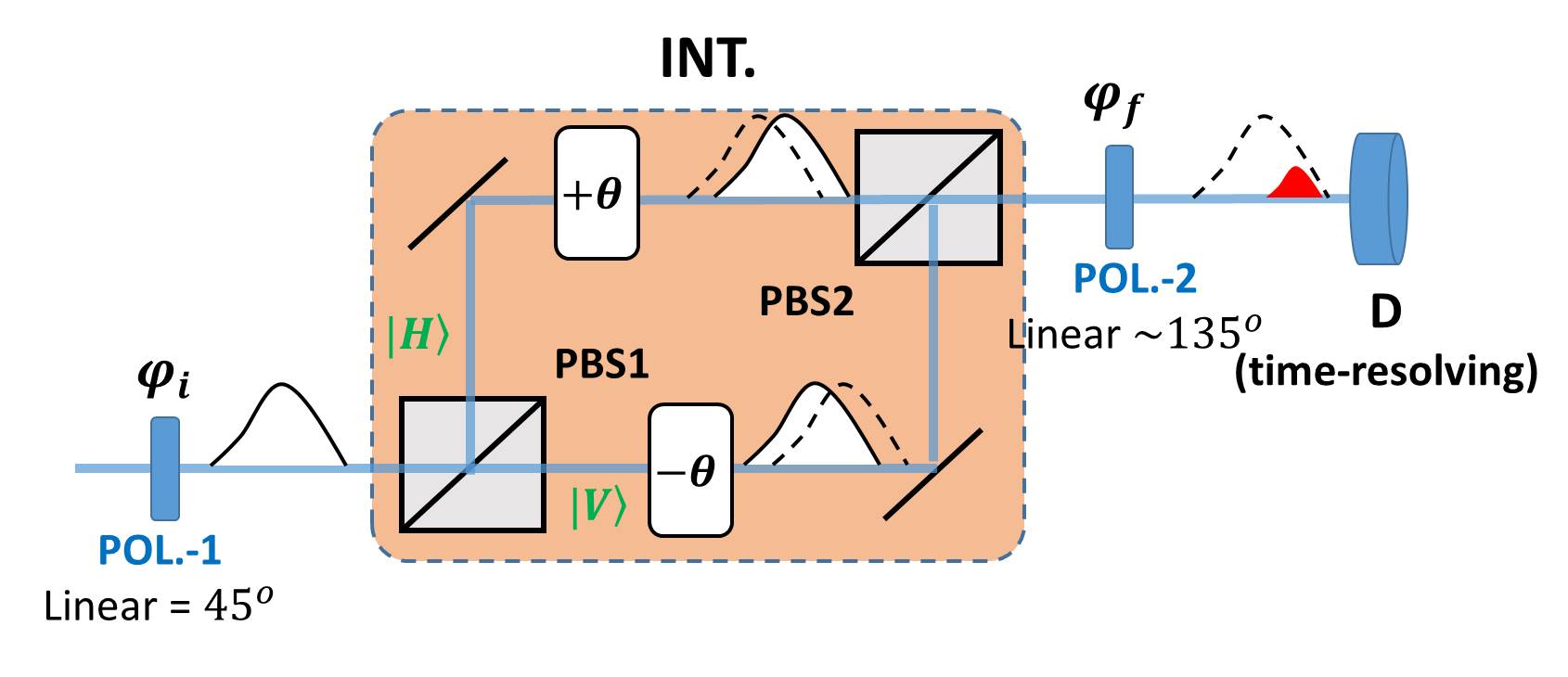}
		\end{minipage}
	}
	\caption{(Color online). The paradigms of previous phase-shift measurement proposals: (a) the standard interferometry scheme and (b) the standard weak measurement scheme. $\pm\theta$: the unknown phase-shift; POL: linear polarizer; PBS: polarizing beam splitter; D: conventional photo-detector for standard interferometry scheme and time-resolving photo-detector for standard weak measurement scheme.}
	\label{fig_scheme_0}
\end{figure}

\section{Robustness of DWM against practical imperfections}
\subsection{Alignment Errors and Detection Errors}
In practice, the accuracy and precision of a phase-shift measurement scheme is not only limited by statistics, but also by the practical imperfections. Here, we consider two important imperfections exist in experiments: alignment error and light intensity variation error, and show that in the present of these imperfections, DWM scheme outperforms standard interferometry (SI) and standard weak measurement (SWM) schemes. 

\textit{Alignment errors}.\textemdash We first consider the alignment error which influence the measurement accuracy. In the DWM scheme, alignment error occurs when the splitting ratio of BS2 (see Fig.\ref{fig_scheme_open}(a)) is deviated from $1:1$ to $(1+\epsilon/2):(1-\epsilon/2)$. Similar to the analysis in Ref.\cite{Brunner2010}, the bias of $\epsilon$ gives potentially erroneous detections in each detector with probability of $\epsilon$, which leads to the requirement of $A_w\theta > \epsilon$. Thus the accuracy limit of DWM scheme is given by:
\begin{equation}\label{eq:re-dwm}
\Delta\theta_{A(DWM)} > \gamma\epsilon.
\end{equation}
By setting $\gamma = 10^{-2}$, we get $\Delta\theta_{A(DWM)} > 10^{-2}\epsilon$. 

We compare the DWM scheme with the standard interferometry (SI) and standard weak measurement (SWM) schemes, which are depicted in Fig.\ref{fig_scheme_0}, and their accuracy limits under alignment error are well analyzed in Ref.\cite{Brunner2010}. For SI, the accuracy limit is
$$\Delta\theta_{A(SI)} > \epsilon,$$
while for SWM, the accuracy limit is given by:
$$\Delta\theta_{A(SWM)} > \omega_0\Delta t\cdot\epsilon,$$
where $\omega_0$ is the central optical frequency of the input light, and $\Delta t$ is the temporal resolution of the time-resolving detector. The current available temporal resolution is around 10ps\cite{Gol2005}, which limits $\Delta\theta_{A(SWM)}$ at the level of $10^4\epsilon$\cite{Brunner2010}.

The analysis above leads us to conclude that in the present of misalignment error $\epsilon$, the accuracy of DWM scheme has improvements of 2 orders and 6 orders of magnitudes comparing to SI scheme and SWM scheme respectively under our settings.

\textit{Detection errors}.\textemdash We now turn to the detection errors, which influence the measurement precision. To be specific, we take two types of errors into account: the error induced by shot noise (denoted as $\Delta I_{SN}$) and the error induced by relative intensity noise (denoted as $\Delta I_{RIN}$). They can be respectively calculated by\cite{Ralph2004}: 
\begin{equation}\label{eq:del}
\begin{array}{lll}
\Delta I_{SN} = \alpha\sqrt{I_D},\\
\Delta I_{RIN} = \beta I_D.
\end{array}
\end{equation}
Here, $\alpha$ and $\beta$ are constants determined by experimental settings, and $I_D$ is the total detected light intensity, which is given by
\begin{equation}\label{eq:I_D_dwm}
I_D = I_{D1} + I_{D2} = \gamma^2I_0.
\end{equation}
Combining Eq.(\ref{eq:dwm}) and Eq.(\ref{eq:del}), the signal-to-noise ratio (SNR) of DWM scheme can be derived as: 
\begin{equation}\label{eq:snr_dwm}
\begin{array}{lll}
SNR_{(DWM)} &= \frac{I_{S(DWM)}}{\Delta I_{SN} + \Delta I_{RIN}} = \frac{2A_w\theta I_D}{\alpha\sqrt{I_D}+\beta I_D}\\
          &= \frac{2\theta}{\alpha\sqrt{1/I_0}+\beta\gamma}.
\end{array}
\end{equation}
To effectively measure the phase-shift, it requires $SNR \geq 1$. The precision limit of DWM is given by:  
\begin{equation}\label{eq:pre_dwm}
\begin{array}{lll}
\Delta\theta_{D(DWM)} \geq \frac{1}{2}(\alpha\sqrt{1/I_0}+\beta\gamma).
\end{array}
\end{equation}

For the standard interferometry scheme, it is easy to verified that the detected intensity and output signal intensity are:
\begin{equation}\label{eq:I_D_si}
I_{D(SI)} = I_0, ~and~ I_{S(SI)} \approx 2 I_0 \theta.
\end{equation}
The SNR for SI is:
\begin{equation}\label{eq:snr_si}
\begin{array}{lll}
SNR_{(SI)} &= \frac{I_{S(SI)}}{\Delta I_{SN} + \Delta I_{RIN}} = \frac{2\theta I_D}{\alpha\sqrt{I_D}+\beta I_D}\\
&= \frac{2\theta}{\alpha\sqrt{1/I_0}+\beta}.
\end{array}
\end{equation}
Therefore the precision limit of SI is given by:
\begin{equation}\label{eq:pre_si}
\begin{array}{lll}
\Delta\theta_{D(SI)} \geq \frac{1}{2}(\alpha\sqrt{1/I_0}+\beta).
\end{array}
\end{equation}

Apparently, the precision limit of DWM is always better than that of SI. Especially, it is useful to consider two extreme situations, in which the photo-detector is hard to saturate and easy to saturate\cite{Harris2017}.

In the first situation, the photo-detector is hard to saturate, so that the input light intensity can be sufficiently large ($I_0 \gg (\frac{\alpha}{\beta})^2$). Applying this condition to Eq.(\ref{eq:snr_dwm}) and Eq.(\ref{eq:snr_si}), we have
\begin{equation}\label{eq:snr_compare}
\Delta\theta_{D(DWM)} \approx \gamma\Delta\theta_{D(SI)}.
\end{equation}

In the second situation, the photo-detector is easy to saturate, and the maximum detectable light intensity is assumed to be $I_{max}$. In this case, the detected light intensity is limited by $I_D < I_{max}$, so that there is an upper bound on the input light intensity. According to Eq.(\ref{eq:I_D_dwm}) and Eq.(\ref{eq:I_D_si}), the upper bounds for DWM and SI are $I_{0(DWM)} < I_{max}/\gamma^2$ and $I_{0(SI)} < I_{max}$ respectively. Inserting these upper bounds into Eq.(\ref{eq:snr_dwm}) and Eq.(\ref{eq:snr_si}), we can also attain $\Delta\theta_{D(DWM)} \approx \gamma\Delta\theta_{D(SI)}$, which is the same as Eq.(\ref{eq:snr_compare}).
Therefore, either the photo-detector is hard to saturate or easy to saturate, the minimum measurable phase-shift of DWM is smaller than SI by a factor of $\gamma$, which is $10^{-2}$ under our settings.

\subsection{Decoherence}\label{sec:imaginary}
In the above analysis, $A_w$ is assumed to be real, and the harmful effect of decoherence is not yet considered.
Interestingly, it has been shown in Ref.\cite{Pang2016} that the imaginary part of weak value can help significantly reducing the systematic errors introduced by decoherence. Inspired by this result, we suggest the possibility of increasing the robustness of DWM under decoherence by modulating a complex weak value.

\begin{figure}[!h]
	\resizebox{7cm}{!}{
		\includegraphics{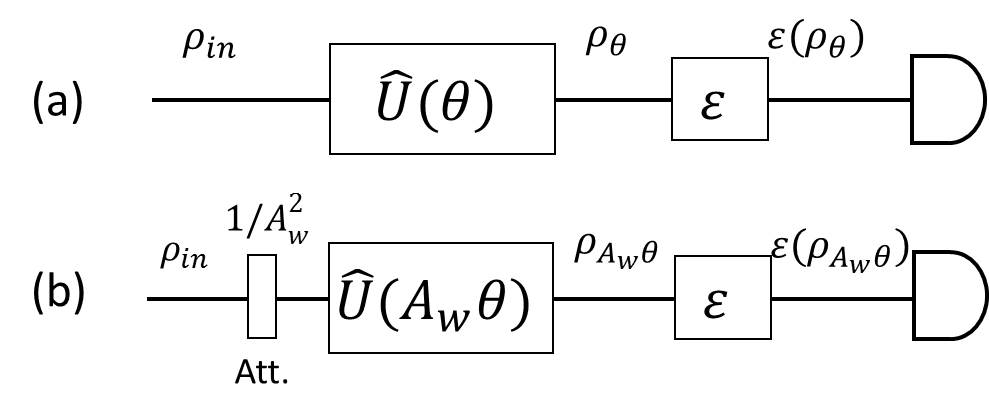}}
	\caption{(Color online). The circuit diagrams describing phase-shift measuring schemes involving decoherence process. (a) The standard interferometry scheme. (b) The difference weak measurement scheme. }\label{fig_circuit}
\end{figure}

In theory, a phase shift measurement using the SI scheme can be described by an input state $\rho_{in}$ experiencing an unitary operation $U(\theta) = \exp[i\theta\hat{Z}]$, where $\hat{Z} = |0\rangle\langle 0| - |1\rangle\langle 1|$ is the Z-Pauli matrix, with $|0\rangle$ and $|1\rangle$ represent the basis. After the unitary operation, the input state becomes $\rho_{\theta}=\hat{U}\rho_{in}\hat{U}^{\dagger}$. A followed decoherence process $\varepsilon$ introduces systematic error, and make output state to be $\varepsilon(\rho_{\theta})$ (see Fig.\ref{fig_circuit}(a)). Similarly, the DWM scheme depicted in Fig.\ref{fig_scheme_open}(b) can be described by an input state experiencing an unitary operation $U(A_w\theta) = \exp[i\theta\hat{Z}]$ with an attenuation factor of $1/A_w^2$ (see Fig.\ref{fig_circuit}(b)). 

%

In our case, the input state $\rho_{in}$ is a two-level state, which can be fully described by a Bloch vector in the Bloch representation\cite{Nielsen2000}: 
$$\rho_{in} = \frac{\hat{I}+\overrightarrow{n}_{in}\cdot\overrightarrow{\hat{\sigma}}}{2},$$
where $\hat{I}$ is the identity matrix, $\overrightarrow{n}_{in}$ is the Bloch vector, and $\overrightarrow{\hat{\sigma}}$ comprises of Pauli matrices in three directions. 
After the phase-shift and decoherence processes (which is denoted as $\varepsilon[\bullet]$), the Bloch vector evolves to:
\begin{equation}
\begin{array}{lll}
SI &: \overrightarrow{n}_{in} \rightarrow \overrightarrow{n}_{\theta} \rightarrow \varepsilon[\overrightarrow{n}_{\theta}],\\
DWM &: \overrightarrow{n}_{in} \rightarrow \overrightarrow{n}_{A_w\theta} \rightarrow \varepsilon[\overrightarrow{n}_{A_w\theta}].
\end{array}\label{eq:n}
\end{equation}
In general, the decoherence process changes the Bloch vector in the following way\cite{Zhong2013}:
$$\varepsilon[\overrightarrow{n}] = E\cdot\overrightarrow{n} + \overrightarrow{C},$$
where $E$ is a $3\times 3$ matrix and $\overrightarrow{C}$ a two-level vector.

To estimate the ultimate measurement precision one can achieve from the output state, we apply the quantum Fisher information (QFI), which can be calculated by\cite{Zhong2013}:
\begin{equation}
F_{\theta} = \left\{\begin{array}{ll}
|\partial_{\theta}\overrightarrow{n}|^2 + \frac{(\overrightarrow{n}\cdot\partial_{\theta}\overrightarrow{n})^2}{1-|\overrightarrow{n}|^2} & \textrm{if $|\overrightarrow{n}|<1$}\\
|\partial_{\theta}\overrightarrow{n}|^2 & \textrm{if $|\overrightarrow{n}|=1$}.
\end{array} \right.\label{eq:fisher}
\end{equation}
Note that when calculating QFI in DWM, the factor of $1/A_w^2$ must be multiplied. By repeating the measuring process for $N$ times, the precision limit can be derived from the Cramer-Rao bound\cite{CRB}:
$$\Delta\theta > \frac{1}{\sqrt{N}F_{\theta}},$$

\begin{figure}[!h]
	\centering
	\subfigure[]{
		\begin{minipage}{4cm}
			\centering
			\includegraphics[width=1.0\textwidth]{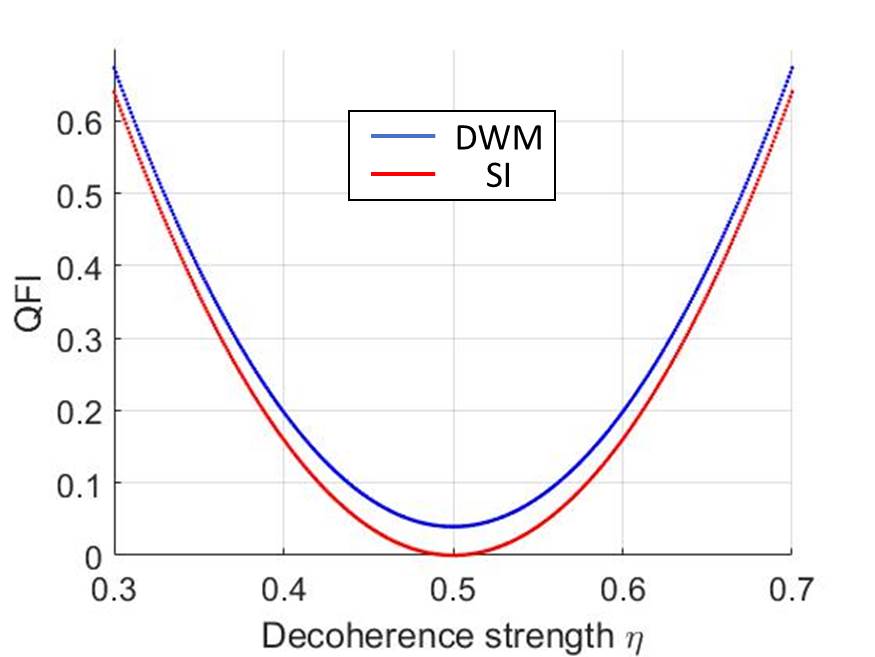}
		\end{minipage}
	}
	\subfigure[]{
		\begin{minipage}{4cm}
			\centering
			\includegraphics[width=1.0\textwidth]{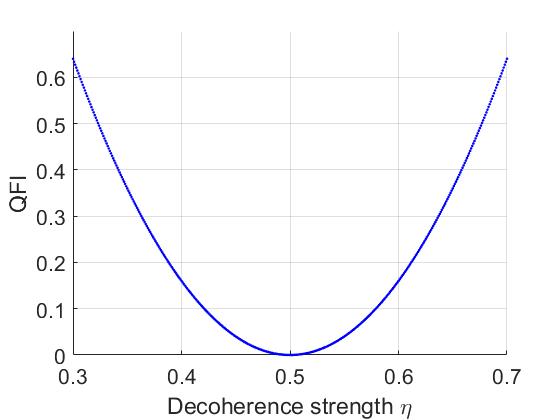}
		\end{minipage}
	}	
	\caption{(Color online) Comparison of quantum Fisher information of DWM (in blue) and SI (in red) under phase flip decoherence, with different decoherence strength of $\eta$. Here we set $\rho_{in} = |+\rangle\langle+|$ and $\theta=10^{-3}$ rad. In (a), the weak value is set to be $A_w = 100 + i10$, and the attainable Fisher information of DWM is higher than that of SI under any decoherence strength. In (b), the weak value is set to be $A_w = 100$, and there is nearly no difference on the attainable Fisher information between DWM and SI.}
	\label{fig_decoherence}
\end{figure}

Combining Eq.(\ref{eq:n}) and Eq.(\ref{eq:fisher}), we are now ready to compare the ultimate precision of DWM and SI under decoherence, using quantum Fisher information as the figure of merit. 
For simplicity, we consider the phase-flip decoherence\cite{Nielsen2000} as an example. The phase-flip decoherence can be described by $E$ and $\overrightarrow{C}$ in the following forms:
$$E_{PF}=\left( \begin{smallmatrix} 1-2\eta & 0 & 0\\ 0 & 1-2\eta & 0 \\ 0 & 0 & 1  \end{smallmatrix} \right), \overrightarrow{C}_{PF} = \left( \begin{smallmatrix} 0\\ 0 \\ 0  \end{smallmatrix} \right),$$ 
where $\eta$ is the decoherence strength. Numerical simulation results are shown in Fig.\ref{fig_decoherence}. Here, the phase-shift and input state are set to be $\theta = 1\times10^{-3}$ rad and $\rho_{in} = |+\rangle\langle+|$ ($|+\rangle \equiv (|0\rangle+|1\rangle)/\sqrt{2}$) respectively. $\Re A_w$ and $\Im A_w$ represent the real part and imaginary part of $A_w$ respectively. When the weak value is complex ($A_w = 100 + i10$), the attainable Fisher information of DWM is always higher than that of SI under any strength of decoherence (see Fig.\ref{fig_decoherence} (a)). 
On the other hand, when the weak value is real ($A_w = 100$), the achievable Fisher information of DWM and SI are almost equal, thus DWM shows no advantage over SI (see Fig.\ref{fig_decoherence} (c)). 

\begin{figure}[!h]
	\centering
	\subfigure[]{
		\begin{minipage}{4cm}
			\centering
			\includegraphics[width=1.0\textwidth]{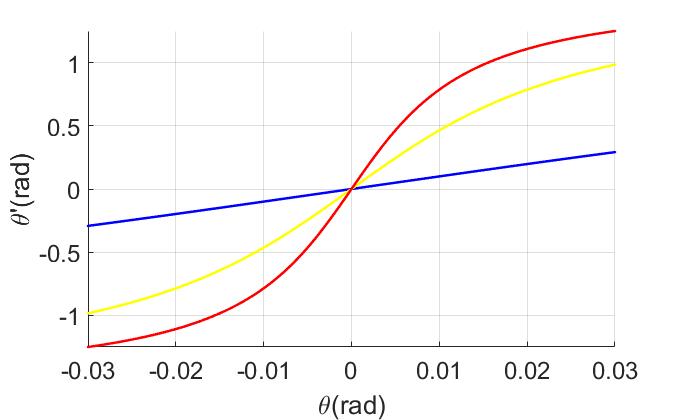}
		\end{minipage}
	}
	\subfigure[]{
		\begin{minipage}{4cm}
			\centering
			\includegraphics[width=1.0\textwidth]{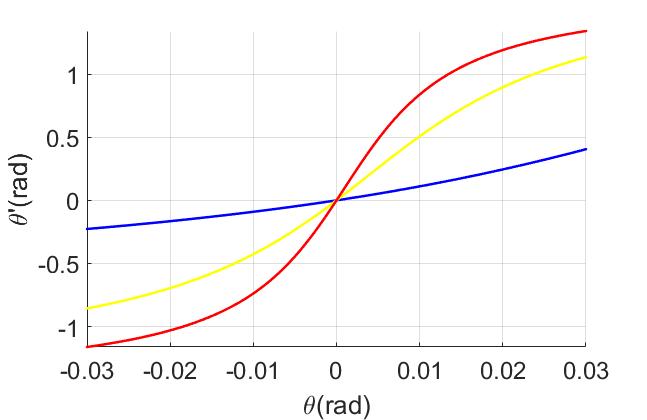}
		\end{minipage}
	}
	\caption{(Color online). Comparison of the phase amplification effects when (a) $\Im A_w = 0$ and (b) $\Im A_w=10$, with $\Re A_w$ equals to: 10 (blue), 50 (yellow) and 100 (red).}
	\label{fig_curve_2}
\end{figure}

Finally, we should note that the imaginary part of weak value would slightly affects the phase amplification factor, as is shown in Fig.(\ref{fig_curve_2}). This deviation can be compensated by data post-processing.

The numerical simulation results imply that it is possible to utilize the imaginary part of $A_w$ for suppressing the effect of decoherence\cite{Pang2016}, and a more rigorous analysis on the general rules behind this phenomenon is worthy for further investigations.

\section{Dynamic Range Extending}\label{sec:closed-loop}
\begin{figure}[!h]
	\resizebox{7cm}{!}{
		\includegraphics{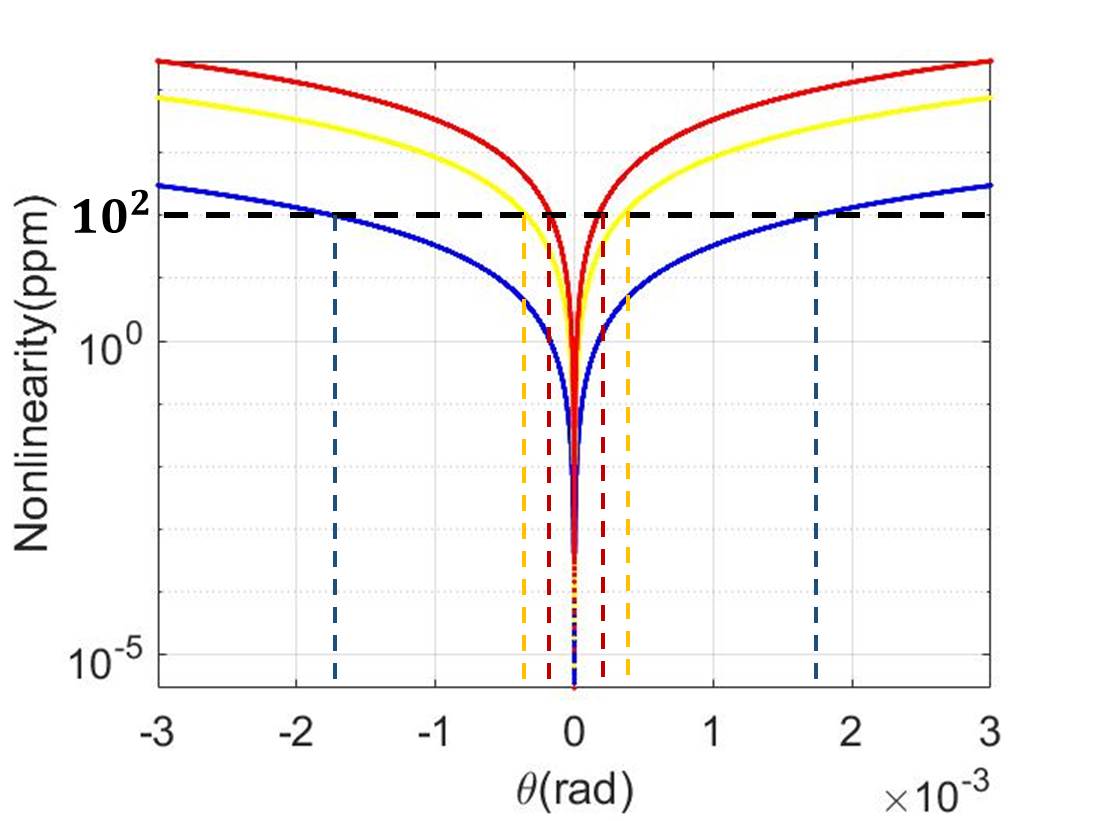}}
	\caption{(Color online). The relation between the nonlinearity (in the unit of parts per million (ppm)) and the phase-shift to be measured. Different colors correspond to different amplification factors ($A_w$): 10(blue), 50(yellow), and 100 (red) respectively. By setting a threshold on nonlinearity, e.g. $10^2$ppm, one can obtain the maximum measurable phase-shifts under different amplification factors.
	}\label{fig_ppm}
\end{figure}

According to Eq.(\ref{eq:dwm}), the phase-shift $\theta$ is approximately linear to the light intensity variation only if $A_w\theta\ll1$ (here we assume that $A_w$ is real for simplicity). When $\theta$ grows up, at some point the effect of nonlinearity can not be neglected. The nonlinearity is given by the deviation between $\theta'$ and $A_w\theta$:
\begin{equation}
D = |1-\frac{\theta'}{A_w\theta}|,
\end{equation}
The relation between $D$ and $\theta$ in DWM is shown in Fig.\ref{fig_ppm}.

To make $D$ sufficiently small, the maximum measurable phase-shift has an upper bounded. For instance, if we need $D < 10^{-4}$, $|\theta'|$ should not be larger than $0.0175$ rad, which requires 
$$|\theta_{max(DWM)}| \approx 0.0175\gamma~rad.$$ 
Note that this upper bound is also available for SWM. As DWM has much lower minimum detectable phase-shift according to the analysis in Sec.III.A, DWM has much higher dynamic range (which is defined as
$R=|\theta_{max}|/|\theta_{min}|$), comparing to SWM.

On the other hand, for SI the maximum measurable phase-shift is $0.0175$ rad, which is larger than DWM by a factor of $1/\gamma$. As the minimum measurable phase-shift of DWM is smaller than SI by a factor of $\gamma$ (see Sec.II.A), the dynamic ranges of DWM and SI are roughly the same. 
%
\begin{figure}[!h]
	\centering
	\subfigure[]{
		\begin{minipage}{8cm}
			\centering
			\includegraphics[width=0.9\textwidth]{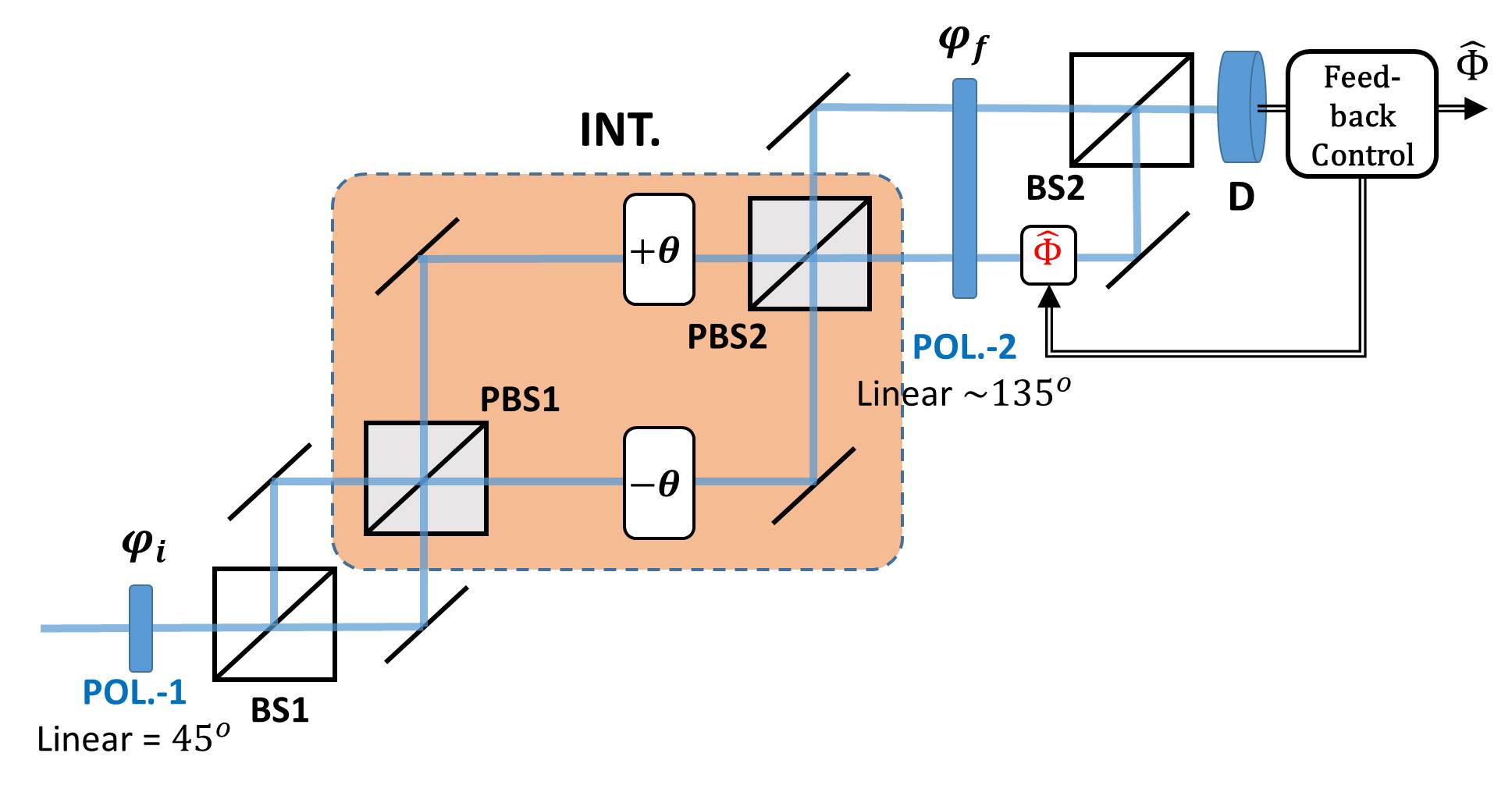}
		\end{minipage}
	}
	\subfigure[]{
		\begin{minipage}{8cm}
			\centering
			\includegraphics[width=0.9\textwidth]{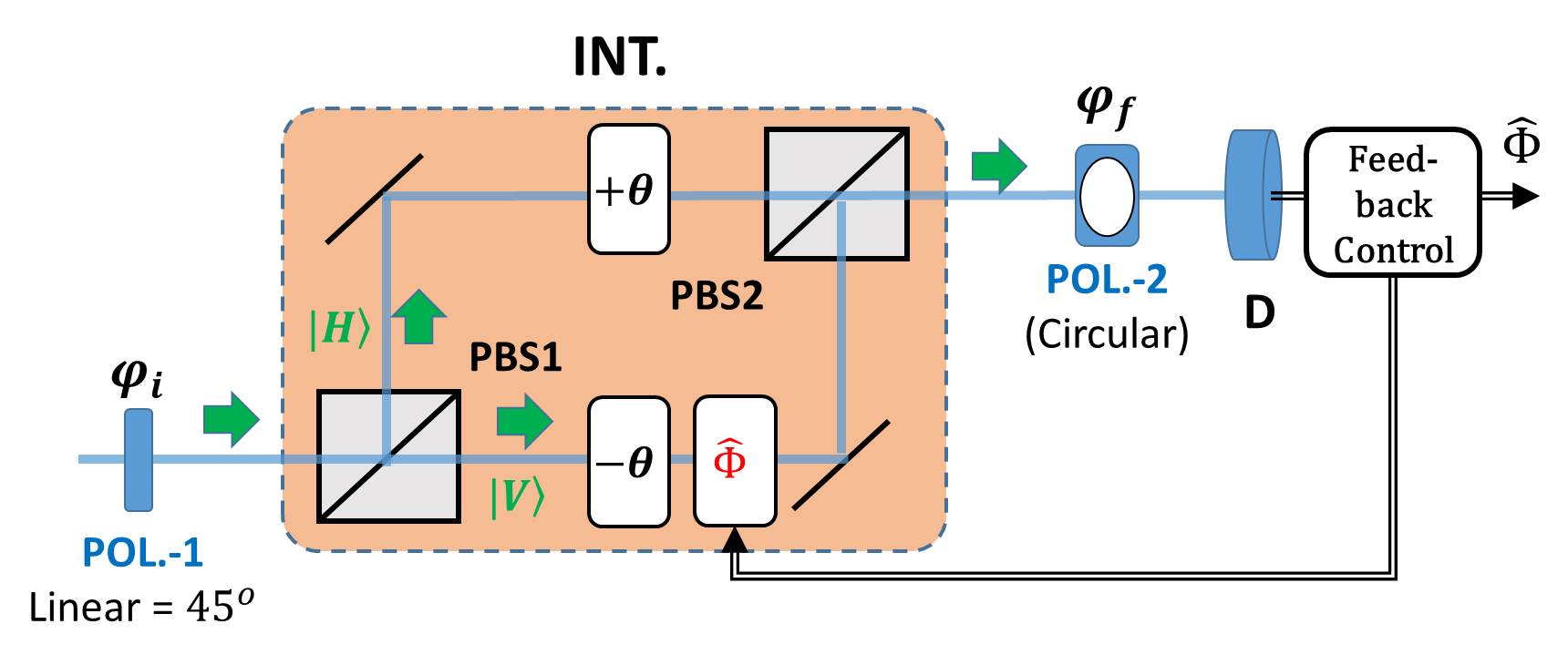}
		\end{minipage}
	}
	\caption{(Color online). (a)The schematic diagram of difference weak measurement in closed-loop scenario. (b) The schematic diagram of standard interferometry in closed-loop scenario. $\pm\theta$: the unknown phase-shift; $\hat{\phi}$: the phase-shift for compensation; POL: linear or circular polarizer; PBS: polarizing beam splitter; D: detector.}
	\label{fig_scheme_closed}
\end{figure}

To overcome this disadvantage and extend the dynamic range of DWM, we propose a closed-loop scenario as is shown in Fig.\ref{fig_scheme_closed}(a). In this scenario, an active modulated phase $\hat{\phi}$ is applied to compensate the phase-shift induced by optical path difference, so that the output from detector is fixed to a certain point. As a comparison, the closed-loop scenario for standard interferometry is depicted in Fig.\ref{fig_scheme_closed}(b). The most important difference between these scenarios is that in DWM it requires 
$$\hat{\phi}_{(DWM)} \simeq \theta/\gamma,$$ 
while in SI it requires 
$$\hat{\phi}_{(SI)} \simeq \theta.$$ 

In practice, suffering from the practical imperfections such as electrical noises, $\hat{\phi}$ can not be modulated in arbitrary precision, and the minimum measurable phase-shift in closed-loop scenario is limited by the modulating precision of $\hat{\phi}$. Suppose the minimum $\hat{\phi}$ that can be modulated is $\hat{\phi}_{min}$, the precision limit of DWM and SI in closed-loop scenario are respectively given by:
\begin{equation}
\begin{array}{lll}
\Delta\theta_{C(DWM)} &> \gamma\hat{\phi}_{min},\\
\Delta\theta_{C(SI)}  &> \hat{\phi}_{min},
\end{array}
\end{equation}
Again, the minimum measurable phase-shift of DWM is smaller than that of SI by a factor of $\gamma$.

On the other hand, the maximum measurable phase-shift for close-loop scenario is determined by the phase modulation device, a range of larger than $2\pi$ rad is achievable. For this reason, the limitation of $A_w\theta \ll 1$ is no longer applied to DWM, the narrow dynamic range problem existing in the current weak measurement schemes has been solved.

\section{Conclusion}
In summary, we propose the difference weak measurement scheme and demonstrate it in the context of phase-shift measurement. We show that in the present of alignment and detection errors, difference weak measurement scheme has much better accuracy and precision comparing to the standard interferometry and standard weak measurement schemes. Weak measurement using real weak value has once been considered less advanced comparing to standard interferometry, because the temporal resolution of photo-detector sets a limitation in the previous proposed scheme\cite{Brunner2010}. Our work shows that this conclusion may be not true in the other schemes: difference weak measurement scheme using real weak vale can outperform standard interferometry under different kinds of practical imperfections. 

Moreover, while the real part of weak value is applied to amplify the phase-shift, an additional imaginary part of weak value may be adopted to reduce the systematic errors induced by decoherence. In this case, the amplification factor would be slightly modified, however this deviation can be compensated by data post-processing. Finally, we state the problem of narrow dynamic range: according to the weak value approximation condition, the maximum measurable phase-shift in the standard weak measurement schemes is strictly limited. The difference weak measurement scheme, on the other hand, can be implemented in closed-loop scenario, thus naturally solves this problem, and achieves a much higher precision comparing to the standard closed-loop interferometry.

\begin{acknowledgments}
This work is supported by National
Natural Science Foundation of China (Grant No. 61701302).
\end{acknowledgments}

\end{document}